\documentclass[nofootinbib,aps,reprint,twocolumn,prl]{revtex4-2}

\usepackage[utf8]{inputenc}

\usepackage{amsmath}
\usepackage{mathrsfs}
\usepackage{amssymb}
\usepackage{bbm}
\usepackage{graphicx}
\usepackage{graphics}
\usepackage[small,compact]{titlesec}
\usepackage{xspace}
\usepackage{colordvi}
\usepackage{color}
\usepackage{units}
\usepackage{stackrel}
\usepackage{hyperref}
\hypersetup{
    colorlinks,
    linkcolor={red!50!black},
    citecolor={blue!50!black},
    urlcolor={blue!80!black}
}

\usepackage[capitalise]{cleveref}

\usepackage{slashed}
\usepackage{orcidlink}

\usepackage{ulem}

\renewcommand{\emph}[1]{\textit{#1}}

\newcommand{\vect}[1]{\vec{\mathbf{#1}}}

\newcommand{\EFT}{$\mathrm{EFT}(\slashed{\pi})$\xspace}

\newcommand{\comment}[1]{}

\newcommand{\jjvHe}{{}^3\mathrm{He}}
\newcommand{\jjvH}{{}^3\mathrm{H}}
\newcommand{\nnlo}{\mathrm{NNLO}}

\newcommand{\GT}{\left<\mathbf{GT}\right>}
\newcommand{\F}{\left<\mathbf{F}\right>}

\makeatletter

\newcommand{\Rmnum}[1]{\expandafter\@slowromancap\romannumeral #1@}
\newcommand{\vast}{\bBigg@{4}}
\newcommand{\Vast}{\bBigg@{5}}
\makeatother

\begin{document}

\title{Tritium $\beta$-decay and Proton-Proton Fusion in Pionless Effective Field Theory}

\author{Ha S. Nguyen\,\orcidlink{0009-0006-1121-8568}}
\email{ha.s.nguyen@duke.edu}
\affiliation{Department of Physics, Box 90305, Duke University, Durham, North Carolina 27708, USA}

\author{Jared Vanasse\,\orcidlink{0000-0001-5593-6971}}
\email{jvanass3@fitchburgstate.edu}
\affiliation{Fitchburg State University, Fitchburg MA 01420, USA
}

\date{\today}

\begin{abstract}
The Gamow-Teller and Fermi  matrix elements, $\GT$ and $\F$, respectively, for tritium $\beta$-decay are calculated to next-to-leading order (NLO) in pionless effective field theory in the absence of Coulomb and isospin violation giving the leading order predictions $\GT_{0}=0.9807$ and $\F_{0}=1$.  Using an experimentally determined value for the tritium-$\beta$ decay GT matrix element, the two-body axial current low energy constant is fixed at NLO yielding $L_{1,A}=6.01\pm2.08$~fm$^{3}$ at the renormalization scale of the physical pion mass, which agrees with predictions based on naive dimensional analysis. Finally, the consequences of Wigner-SU(4) spin-isospin symmetry are considered for the Gamow-Teller matrix element.
\end{abstract}


\maketitle


The simplest nuclear system that offers an experimentally clean probe of the axial current is tritium $\beta$-decay.  Due to the small difference in the $\jjvH$-$\jjvHe$ binding energies ($\sim$764~keV), tritium $\beta$-decay near threshold can be described quite naturally in pionless effective field theory (\EFT), which has been used to great success in the description of two- and three-body nuclear systems (See Refs.~\cite{Beane:2000fx,Vanasse:2016jtc} for reviews).  Kong and Ravndal~\cite{Kong:2000px,Butler:2001jj} used \EFT to calculate the  $pp\to de^{+}\nu_{e}$ cross section to next-to-leading order (NLO), but were constrained by the unknown two-body axial current low energy constant (LEC) $L_{1,A}$.  To next-to-next-to-leading order ($\nnlo$), Ref.~\cite{Butler:2000zp} showed that $\nu_{x} d\to np\nu_{x}$ and $\bar{\nu}_{e} d\to nne^{+}$ were also limited by the uncertainty in $L_{1,A}$.  Precise knowledge of these processes is important for determining the total flux of neutrinos from the Sun in the Sudbury Neutrino Observatory experiment.  $L_{1,A}$ also appears in the process $\mu^{-}d\to nn\nu_{\mu}$, which will be measured in the upcoming MuSUN experiment~\cite{Andreev:2010wd,Kammel:2021jss}.

Tritium $\beta$-decay is a super-allowed process that is predominantly given by the Gamow-Teller (GT) and Fermi (F) matrix elements. It is also of inherent interest as a detailed analysis of the tail end of its spectrum can give the mass of the antineutrino~\cite{Drexlin:2013lha} and potentially give evidence for sterile neutrinos~\cite{Mertens:2014nha,Barry:2014ika,KATRIN:2022ayy}.

Using the formalism of Ref.~\cite{Vanasse:2017kgh}, this letter calculates the GT and F matrix elements of tritium-$\beta$ decay to NLO in \EFT.  Isospin invariance is assumed and Coulomb interactions are neglected but their contribution is estimated to be small.  Such effects can be treated as perturbative corrections and making these assumptions removes the need for an additional Coulomb dependent isospin breaking three-body force correction~\cite{Vanasse:2014kxa,Konig:2015aka}.  The experimentally determined GT matrix element is used to fit $L_{1,A}$ at NLO.  In addition, the consequences of Wigner-SU(4)\,spin-isospin symmetry on the GT matrix element are explored and the order at which three-nucleon effects will become important for the axial current is discussed.



\textit{Lagrangian and Two-Body System:} The Lagrangian for \EFT including the weak axial and vector currents in the dibaryon formalism is given by
\begin{align}
&\mathcal{L}=\hat{N}^{\dagger}\left(i\partial_{0}+\frac{\vect{\nabla}^{2}}{2M_{N}}\right)\hat{N}+\frac{g_{A}}{\sqrt{2}}\hat{N}^{\dagger}\sigma_{i}\tau^{+}\hat{N}\hat{A}_{i}^{-}\\\nonumber
&+g_{V}\hat{N}^{\dagger}\tau^{+}\hat{N}\hat{V}_0^{-}\\\nonumber
&+\hat{t}_{i}^{\dagger}\left[\Delta_{t}-c_{0t}\left(i\partial_{0}+\frac{\vect{\nabla}^{2}}{4M_{N}}+\frac{\gamma_{t}^{2}}{M_{N}}\right)\right]\hat{t}_{i}\\\nonumber
&+\hat{s}_{a}^{\dagger}\left[\Delta_{s}-c_{0s}\left(i\partial_{0}+\frac{\vect{\nabla}^{2}}{4M_{N}}+\frac{\gamma_{s}^{2}}{M_{N}}\right)\right]\hat{s}_{a}\\\nonumber
&+y\left[\hat{t}_{i}^{\dagger}\hat{N} ^{T}P_{i}\hat{N} +\hat{s}_{a}^{\dagger}\hat{N}^{T}\bar{P}_{a}\hat{N}+\mathrm{H.c.}\right]\\\nonumber
&+\hat{\psi}^{\dagger}\Omega\hat{\psi}+\sum_{n=0}^{1}\left[\omega^{(n)}_{t0}\hat{\psi}^{\dagger}\sigma_{i}\hat{N}\hat{t}_{i}
-\omega^{(n)}_{s0}\hat{\psi}^{\dagger}\tau_{a}\hat{N}\hat{s}_{a}+\mathrm{H.c.}\right]\\\nonumber
&+l_{1,A}\hat{t}_{k}^{\dagger}\hat{s}_{-}\hat{A}_{k}^{-}++l_{1,V}\hat{s}_{3}^{\dagger}\hat{s}_{-}\hat{V}_0^{-}+\mathrm{H.c.},
\end{align}
where $\hat{t}_{i}$ ($\hat{s}_{a}$) is the spin-triplet (spin-singlet) dibaryon, $\hat{\psi}$ is the iso-doublet three-nucleon field, $\hat{A}_{k}^{-}$ is the leptonic axial current, and $\hat{V}_0^{-}$ is a component of the leptonic vector current.  The strong interaction parameters are fit using the $Z$-parametrization~\cite{Phillips:1999hh}, which at leading order (LO) fits to the nucleon-nucleon ($N\!N$) scattering $^{3}\!S_{1}$ and $^{1}\!S_{0}$ poles and their residues at NLO, yielding
\begin{align}
y^{2}=\frac{4\pi}{M_{N}},\, &\Delta_{t}=\gamma_{t}-\mu,\, c_{0t}^{(n)}=(-1)^{n}(Z_{t}-1)^{n+1}\frac{M_{N}}{2\gamma_{t}},\\\nonumber
&\Delta_{s}=\gamma_{s}-\mu,\, \!c_{0s}^{(n)}=(-1)^{n}(Z_{s}-1)^{n+1}\frac{M_{N}}{2\gamma_{s}}.
\end{align}
where $\gamma_{t}=45.7025$~MeV ($\gamma_{s}=-7.890$~MeV) is the $^{3}\!S_{1}$ bound state ($^{1}\!S_{0}$ virtual bound state) momentum and $Z_{t}=1.6908$ ($Z_{s}=0.9015$) is the residue about the $^{3}\!S_{1}$ ($^{1}\!S_{0}$) pole.  Three-body parameters $\Omega$ and $\omega_{\{t,s\}0}^{(n)}$ are fit to the $\jjvH$ binding energy~\cite{Vanasse:2015fph} and $P_{i}=\frac{1}{\sqrt{8}}\sigma_{2}\sigma_{i}\tau_{2}$ ($\bar{P}_{a}=\frac{1}{\sqrt{8}}\tau_{2}\tau_{a}\sigma_{2}$) projects out the spin-triplet iso-singlet (spin-singlet iso-triplet) combination of nucleons..  The Pauli matrix $\tau^{+}$ is normalized such that $\tau^{+}=-(\tau_{1}+i\tau_{2})/\sqrt{2}$ and the one-nucleon axial (vector) coupling is $g_{A}=1.26$ ($g_{V}=1$).  The two-body axial current LEC $l_{1,A}$ in the dibaryon formalism is related to the traditional $L_{1,A}$ coupling in Ref.~\cite{Butler:1999sv} via %
\begin{align}
\label{eq:matching}
&l_{1,A}=-\frac{M_{N}}{4\pi}\Delta_{t}\Delta_{s}L_{1,A}+\frac{g_{A}}{{2}}\left(c_{0t}^{(0)}\frac{\Delta_{s}}{\Delta_{t}}+c_{0s}^{(0)}\frac{\Delta_{t}}{\Delta_{s}}\right)\\\nonumber
&l_{1,V}=g_{V}c_{0s}^{(0)}.
\end{align}
The additional terms in $l_{1,A}$ and $l_{1,V}$ are induced by the coordinate transformation relating the nucleon formalism to the dibaryon formalism.  $l_{1,V}$ is entirely predicted by other known LECs and contains no new two-body vector current LEC.\newline



\textit{GT and Fermi Matrix Elements:} The half life of $\jjvH$ $\beta$-decay $t_{1/2}$ is given by~\cite{Schiavilla:1998je} 
\begin{equation}
\frac{(1+\delta_{R})f_{V}}{K/G_{V}^{2}}t_{1/2}=\frac{1}{\left<\mathbf{F}\right>^{2}+f_{A}/f_{V}g_{A}^{2}\left<\mathbf{GT}\right>^{2}}
\end{equation}
where $\F$ ($\GT$) is the Fermi (Gamow-Teller) matrix element and other parameters are given in Ref.~\cite{Schiavilla:1998je}.

\begin{figure}[hbt]
\includegraphics[width=80mm]{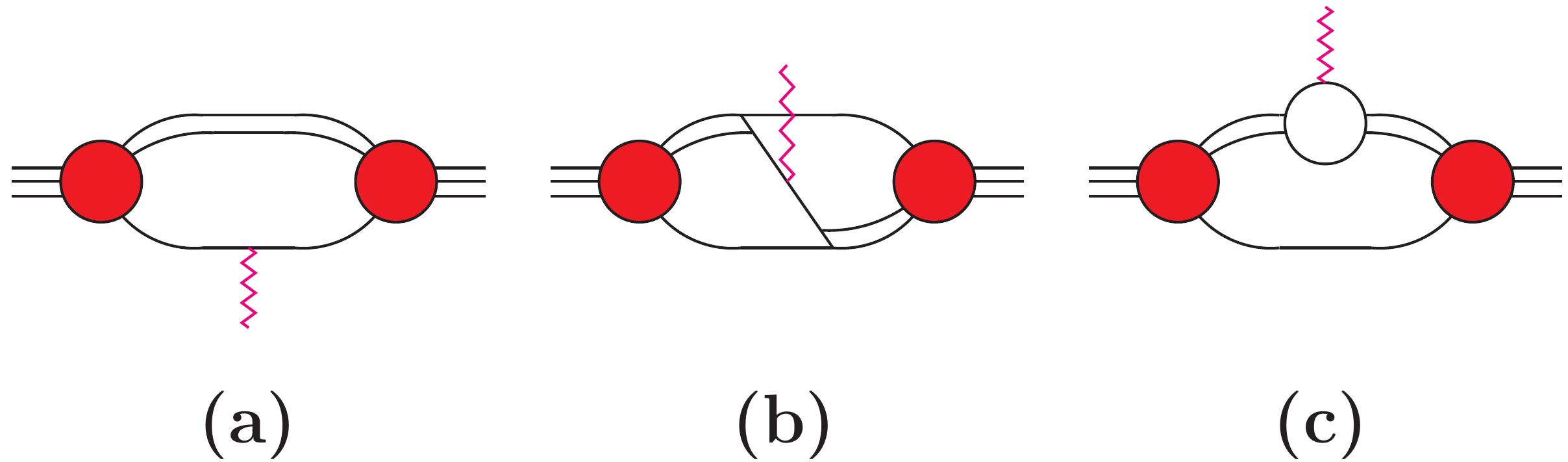}
\caption{Diagrams for the LO three-nucleon axial and vector form factors.  The zig-zag represents the one-body axial or vector current.\label{fig:FormFactorLO}}
\end{figure}

In the absence of Coulomb interactions and assuming isospin invariance, the axial and vector form factor can be calculated using Ref.~\cite{Vanasse:2017kgh}.  In the limit $Q^{2}=0$, the axial (vector) form factor gives the GT (F) matrix element.  The axial (vector) form factor is given by the sum of diagrams in Fig.~\ref{fig:FormFactorLO}
where single lines are nucleons, double lines dibaryons, triple lines three-nucleon systems, circles three-nucleon vertex functions, and zig-zag lines the axial or vector current.  Details of how these diagrams and the three-nucleon vertex functions are calculated can be found in Refs.~\cite{Vanasse:2015fph,Vanasse:2017kgh}.  The LO GT (F) matrix element is given by
\begin{align}
\label{eq:LOGT}
F_{W,0}^{\chi}(Q^2=0)=2\pi M_{N}\left(\widetilde{\boldsymbol{\Gamma}}_{0}(q)\right)^{T}\otimes f_{0}(q,\ell)\otimes\widetilde{\boldsymbol{\Gamma}}_{0}(\ell),
\end{align}
where
\begin{align}
\label{eq:LOGenForm}
&f_{0}(q,\ell)=2\pi M_{N}\left\{g(q,\ell)
\left(\begin{array}{cc}
c_{11}+a_{11} & c_{12} \\
c_{21} & c_{22}+a_{22}
\end{array}\right)\right.\\[2mm]\nonumber
&\hspace{.5cm}\left.+h(q,\ell)
\left(\!\!\begin{array}{cc}
b_{11}-2a_{11} & b_{12}+3(a_{11}+a_{22}) \\
b_{21}+3(a_{11}+a_{22}) & b_{22}-2a_{22}
\end{array}\!\right)\right\},
\end{align}
\begin{equation}
g(q,\ell)=\frac{\pi}{2}\frac{\delta(q-\ell)}{q^{2}\sqrt{\frac{3}{4}q^{2}-M_{N}B}},
\end{equation}
and
\begin{equation}
h(q,\ell)=\frac{1}{q^{2}\ell^{2}-(q^{2}+\ell^{2}-M_{N}B)^{2}}.
\end{equation}
$\chi$ is GT (F) for the axial (vector) form factor.  The function $\widetilde{\boldsymbol{\Gamma}}_{0}(q)$ is related to the three-nucleon vertex function~\cite{Vanasse:2015fph}.  Coefficients $a_{11}$, $b_{11}$, $c_{11}$, and etc. in Eq.~(\ref{eq:LOGenForm}) come from projecting out the one-body axial or vector current for each of the diagrams in the doublet $S$-wave channel giving the values in Table~\ref{tab:LOvalues}. $B=8.48$~MeV is the triton binding energy.

\begin{table}
\begin{tabular}{|c|c|c|c|c|c|c|c|c|c|c|}
\hline
Form factor & $a_{11}$ & $a_{22}$ & $b_{11}$ & $b_{12}$ & $b_{21}$ & $b_{22}$ & $c_{11}$ & $c_{12}$ & $c_{21}$ & $c_{22}$ \\
\hline
$F_{W}^{\mathrm{GT}}(Q^{2})$ & $-\frac{1}{3}$ & $-\frac{1}{3}$ & $\frac{5}{3}$ & $\frac{1}{3}$ & $\frac{1}{3}$ & $\frac{5}{3}$ & 0 & $\frac{2}{3}$ & $\frac{2}{3}$ & 0  \\\hline
$F_{W}^{\mathrm{F}}(Q^{2})$ & 1 & $-\frac{1}{3}$ & 1 & 1 & 1 & -$\frac{5}{3}$ & 0 & 0 & 0 & $\frac{4}{3}$ \\\hline
\end{tabular}
\caption{\label{tab:LOvalues}Values of coefficients for the LO weak form factor used in Eq.\eqref{eq:LOGenForm}}
\end{table}

\begin{figure}[hbt]
\includegraphics[width=80mm]{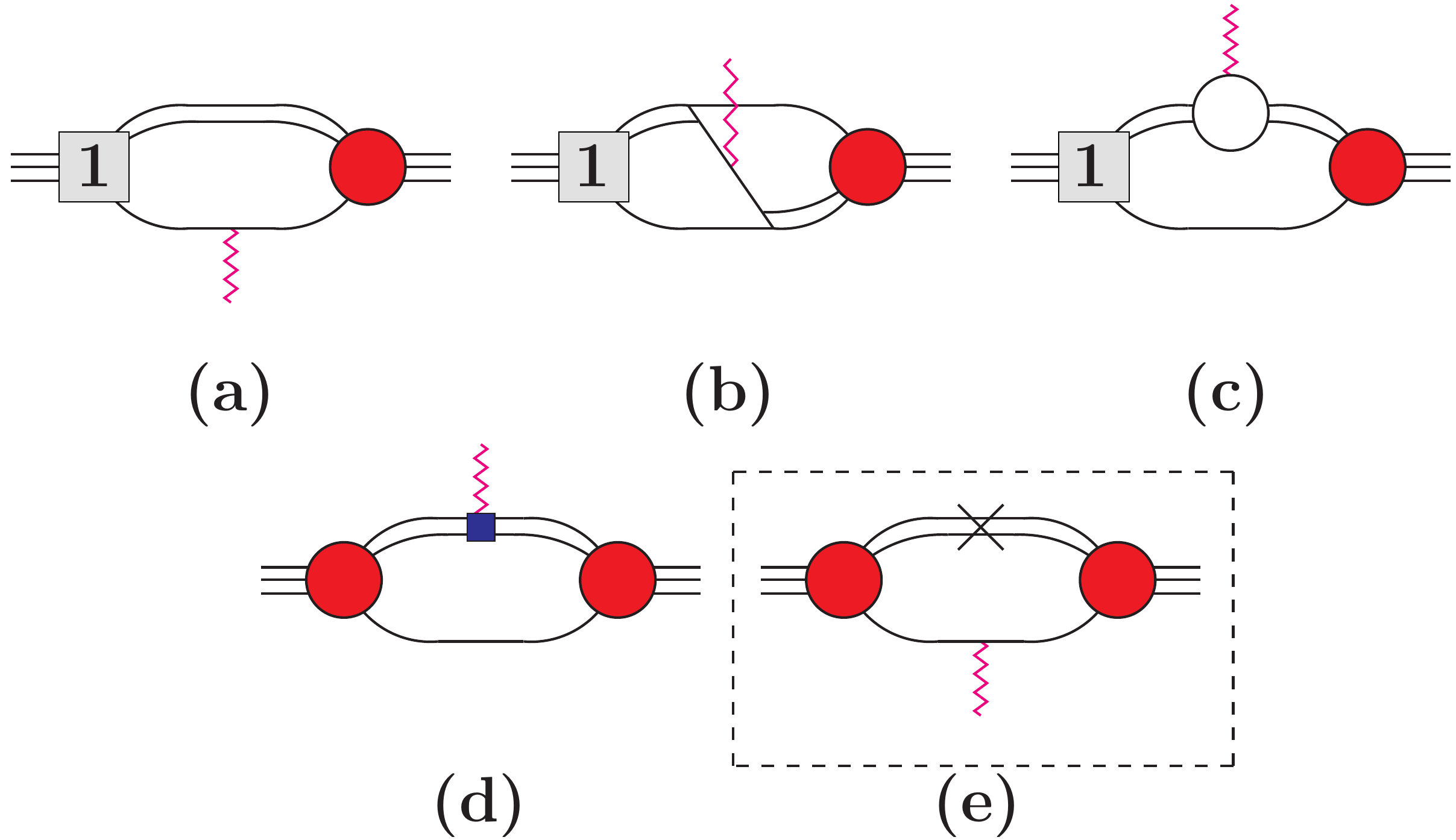}
\caption{NLO corrections to the axial and vector form factor.  The dashed-boxed diagram is subtracted to avoid double counting and the box with 1 is the NLO correction to the three-nucleon vertex function.  Diagram-(d) for the axial (vector) form factor comes from the two-body axial (vector) current term $l_{1,A}$ ($l_{1,V}$).  Diagrams related by time reversal symmetry are not shown.\label{fig:FormFactorNLO}}
\end{figure}

The NLO correction to the axial (vector) form factor is given by the diagrams in Fig.~\ref{fig:FormFactorNLO}
where diagram-(d) is the $l_{1,A}$ ($l_{1,V}$) term for the axial (vector) form factor and diagram-(e) is subtracted to avoid double counting from diagram-(a) and its time reversed version.  In the limit $Q^{2}=0$, these contributions give the NLO correction to the GT (F) matrix element given by 

\begin{align}
\label{eq:NLOform}
&F_{W,1}^{\chi}(Q^2=0)=2\pi M_{N}\left(\widetilde{\boldsymbol{\Gamma}}_{1}(q)\right)^{T}\otimes f_{0}(q,\ell)\otimes\widetilde{\boldsymbol{\Gamma}}_{0}(\ell)\\\nonumber
&\hspace{.5cm}+2\pi M_{N}\left(\widetilde{\boldsymbol{\Gamma}}_{0}(q)\right)^{T}\otimes f_{0}(q,\ell)\otimes\widetilde{\boldsymbol{\Gamma}}_{1}(\ell)\\\nonumber
&\hspace{.5cm}-4\pi M_{N}\left(\widetilde{\boldsymbol{\Gamma}}_{0}(q)\right)^{T}\otimes f_{1}(q,\ell)\otimes\widetilde{\boldsymbol{\Gamma}}_{0}(\ell),
\end{align}
where
\begin{equation}
\label{eq:NLOd}
f_{1}(q,\ell)=\frac{\pi}{2}\frac{\delta(q-\ell)}{q^{2}}
\left(\begin{array}{cc}
\frac{c_{0t}^{(0)}}{M_{N}}a_{11}+d_{11} & d_{12} \\
d_{21} & \frac{c_{0s}^{(0)}}{M_{N}}a_{22}+d_{22}
\end{array}\right).
\end{equation}

Coefficients $d_{11}$,$d_{12}$,$d_{21}$, and $d_{22}$ are given in Table~\ref{tab:dvalues}.
\begin{table}
\begin{tabular}{|c|c|c|c|c|}
\hline
Form factor & $d_{11}$ & $d_{12}$ & $d_{21}$ & $d_{22}$  \\\hline
$F^{GT}_{W}(Q^{2})$ & 0 & $\frac{1}{3}l_{1,A}/M_N$ & $\frac{1}{3}l_{1,A}/M_N$ & 0  \\\hline
$F^{F}_{W}(Q^{2})$ & 0 & 0 & 0 & $\frac{4}{3}c_{0s}^{(0)}/M_{N}$  \\\hline
\end{tabular}
\caption{Values of coefficients for the NLO correction to the axial (vector) form factor used in Eq.~\eqref{eq:NLOd}\label{tab:dvalues}}
\end{table}
$\widetilde{\boldsymbol{\Gamma}}_{1}(q)$ is related to the NLO correction to the three-nucleon vertex function~\cite{Vanasse:2015fph}.
%
%
%
\textit{Results:} The LO GT matrix element is $\GT_{0}=\sqrt{3}\times0.9807$, while the NLO GT matrix element depends on $L_{1,A}$.\footnote{The factor of $\sqrt{3}$ comes from a Clebsch-Gordan coefficient.}  Taking $L_{1,A}=0$ gives $\GT_{0+1}=\sqrt{3}\times0.9935$.  Choosing the renormalization scale $\mu=m_\pi$, a fit of $L_{1,A}$ to the value extracted from experiment $\GT=\sqrt{3}\times0.9511(13)$~\cite{Baroni:2016xll} yields
%
%
\begin{equation}
\label{eq:L1Apred}
L_{1,A}(\mu=m_\pi)=6.01\pm2.08~\mathrm{fm}^{3},
\end{equation}
This is compatible with naturalness expectations, which predicts a value of~\cite{Butler:2001jj}
\begin{equation}
|L_{1,A}(\mu=m_\pi)|\approx\frac{1}{m_{\pi}(m_{\pi}-\gamma_{t})^{2}}=6.5~\mathrm{fm}^{3}
\end{equation}

The $pp$-fusion rate is given by the matrix element
\begin{equation}
 \left|\langle d;j\left|A_{k}^{-}\right|pp\rangle\right|=g_{A}C_{\eta}\sqrt{\frac{32\pi}{\gamma_{t}^{3}}}\Lambda(p)\delta_{k}^{j}
 \end{equation} 
 where $C_{\eta}$ is the Somerfeld factor in Coulomb scattering and $\Lambda(p)$ at threshold to NLO in the $Z$-parametrization is given by~\cite{Butler:2001jj}\footnote{$pp$ fusion in the $Z$-parametrization can be obtained from calculations in the ERE parametrization by replacing all occurences of the effective range $\rho_{t}$ with $(Z_{t}-1)/\gamma_{t}$.  Differences between ERE and $Z$ parametrization in the $^{1}S_{0}$ channel are $\sim$1\% effects~\cite{Griesshammer:2004pe} and can be neglected at NLO.}
\begin{align}
\Lambda(0)&=\frac{1}{2}(1+Z_{t})\left\{e^{\eta}-\gamma_{t}a_{pp}[1-\eta e^{\eta}\Gamma(0,\eta))\right\}\\\nonumber
&-\gamma_{t}^{2}a_{pp}\frac{\gamma_t-\mu}{M_NC_{0,-1}^{(pp)}}\left[\frac{L_{1,A}}{g_A}-\frac{M_N}{2}\left(C_{2,-2}^{(pp)}+C_{2,-2}^{(d)}\right)\right]
\end{align}
The value $\eta=\alpha M_{n}/\gamma_{t}$, $\alpha$ is the fine structure constant from QED, and $\Gamma(0,\eta)$ an incomplete gamma function all arising from Coulomb corrections.  The LECs $C_{0,-1}^{(pp)}$, $C_{2,-2}^{(pp)}$, and $C_{2,-2}^{(d)}$ are given in Ref.~\cite{Butler:2001jj}.  Plugging in physical values, the NLO $Z$-parametrization prediction for $\Lambda(0)$ is
\begin{equation}
\Lambda(0)=2.72+0.0087\left(\frac{L_{1,A}}{1\,\mathrm{fm}^{3}}\right)+\mathcal{O}(12\%)
\end{equation}
Using the value for $L_{1,A}$ from Eq.~(\ref{eq:L1Apred}) gives $\Lambda(0)=2.77(33)$. Within errors, this prediction agrees with the phenomenological value $\Lambda(0)=2.65(1)$~\cite{Adelberger:2010qa}.  

In the isospin limit ignoring higher partial waves, the GT matrix element is given by
\begin{equation}
\left<\mathbf{GT}\right>=\sqrt{3}\times(P_{S}-P_{S'}/3),
\end{equation}
where $P_{S}$ is the probability of the triton wavefunction being in the symmetric $S$ state and $P_{S'}$ is the probability of the mixed symmetry $S'$ state~\cite{Schiavilla:1998je}.  In the  Wigner-SU(4) limit ($\gamma_{t}=\gamma_{s}$, $Z_{t}=Z_{s}$), $P_{S'}=0$ and therefore $\left<\mathbf{GT}\right>=\sqrt{3}$ up to NLO, which is verified numerically.  In order to get $\GT=\sqrt{3}$ at NLO, $l_{1,A}$ is defined by Eq.~(\ref{eq:matching}) with $L_{1,A}=0$.  In the Wigner limit, the non $L_{1,A}$ term in Eq.~(\ref{eq:matching}) becomes $\mu$ independent.  Similarly, since there is no isospin breaking, it is found at LO and NLO that the F matrix element reproduces the wavefunction renormalization expression and therefore $\F_{0}=1$ and $\F_{0+1}=1$, in agreement with the Ademollo-Gatto theorem~\cite{Ademollo:1964sr}.

The GT matrix element has been calculated previously in \EFT with the inclusion of Couloumb interactions in Ref.~\cite{De-Leon:2016wyu}. When they dropped Coulomb and isospin breaking terms, they found $\GT_0=\sqrt{3}$.  However, this result is only true in the Wigner-SU(4) limit and Ref.~\cite{De-Leon:2016wyu} did not specify that they were in the Wigner-SU(4) limit.



 
\textit{Conclusions:} In this work the Gamow-Teller ($\GT$) and Fermi ($\F$)  matrix elements of $\jjvH$ $\beta$ decay have been calculated to NLO in \EFT in the $Z$-parametrization ignoring Coulomb and isospin effects.  The omitted Coulomb effects are perturbative corrections approximately of the size $\alpha M_{n}/p_{*}\sim6\%$, where $p_{*}=\sqrt{(4/3)(M_{N}(B_{\jjvH}+\gamma_{t}^{2})}\sim112$~MeV is the approximate three-nucleon binding momentum.\footnote{This estimate for the binding momentum differs from previous estimates that did not include the binding of the deuteron~\cite{Vanasse:2016umz}.   Previous estimates found Coulomb effects to be slightly larger.}  Coulomb corrections can be included perturbatively as in Ref.~\cite{Konig:2015aka} or in a nonperturbative fashion as in Refs.~\cite{Rupak:2001ci,Vanasse:2014kxa}.  At NLO, the two-body axial current LEC $L_{1,A}=6.01\pm2.08$~fm$^3$ was fit to reproduce the experimentally determined $\GT$ matrix element.  In addition, it was also found that in the Wigner-$\mathrm{SU}(4)$ limit, $\GT=\sqrt{3}$ at LO and NLO (with values for $l_{1,A}$ solely predicted from two-body physics), in agreement with analytical predictions and is a nontrivial check on the calculation.  The value for $\F$ at LO and NLO was found to be 1, which is expected due to the lack of isospin breaking up to NLO and the Adellamo Gatto theorem~\cite{Ademollo:1964sr}.  Finally, using the value for $L_{1,A}$ determined from $\jjvH$ $\beta$ decay, the threshold value for the $pp$-fusion reduced matrix element is $\Lambda(0)=2.77(33)$, with a $12\%$ NLO \EFT error estimate.

\textit{Outlook:} $L_{1,A}$ is the only unknown two-body axial current LEC up to $\nnlo$ in \EFT.  Thus in principle with a prediction of $L_{1,A}$, the $pp$-fusion cross-section could be determined to $~3\%$ with a NNLO \EFT prediction.  However, our predicted value of $L_{1,A}$ relied on a fit to $\jjvH$ $\beta$ decay at NLO in \EFT.  A NNLO \EFT $L_{1,A}$ calculation of $\jjvH$ $\beta$ decay would necessitate refitting $L_{1,A}$ or adding a perturbative correction.  Reference~\cite{Lin:2022yaf} demonstrated that a NNLO \EFT calculation of the three-nucleon magnetic moments requires the insertion of a new three-body current counterterm.  This would imply there is a three-body axial current counterterm.  Therefore, a NNLO calculation of $\jjvH$ $\beta$ decay is not possible without fitting this new three-body axial current counterterm to a new three-body datum.  In the case of $\chi$EFT this would manifest as a three-nucleon meson exhange current that would give a sizable contribution at low energies.  This also implies that any calculation including meson exchange must include three-body meson exhange currents at low energies to make accurate comparisons between $pp$-fusion and $\jjvH$ $\beta$-decay.

\acknowledgments{We would like to thank Roxanne Springer, Xincheng Lin, Brian Tiburzi, and Daniel Phillips for useful discussions during the course of this work.  This material is based upon work supported by the U.S. Department of Energy, Office of Science, Office of Nuclear Physics, under Award Number DE-FG02-93ER40756 (JV) and DE-FG02-05ER41368 (HSN).}

\appendix


\end{document}